\documentclass[eqsecnum,floatfix,aps,prd,floatfix,tightenlines]{revtex4-2} 

\usepackage{graphicx,xcolor,soul,todonotes}
\usepackage{graphics}
\usepackage{bm}
\usepackage{amssymb}
\usepackage{amsmath}
\usepackage{mathrsfs}

\newcommand{\Prob}{\textrm{Prob}}

\begin{document}
\title{Probability Distribution for Vacuum Energy Flux Fluctuations in Two Spacetime Dimensions}

 \author{Christopher J. Fewster}
\email{chris.fewster@york.ac.uk}
\affiliation{Department of Mathematics, University of York, 
Heslington, York YO10 5DD,
United Kingdom}
\affiliation{York Centre for Quantum Technologies, University of York, Heslington, York YO10 5DD, United Kingdom.}

 \author{L. H. Ford}
 \email{ford@cosmos.phy.tufts.edu}
 \affiliation{Institute of Cosmology, Department of Physics and Astronomy, Tufts University, Medford, Massachusetts 02155, USA}
     
     \begin{abstract}
     The probability distribution for vacuum fluctuations of the energy flux in two dimensions will be constructed, along with the joint distribution of energy flux and energy density. Our approach will be based on previous
     work on probability distributions for the energy density in two dimensional conformal field theory. In both cases, the relevant stress tensor component
     must be averaged in time, and the results are sensitive to the form of the averaging function. Here we present results for two classes of such functions,
     which include the Gaussian and Lorentzian functions.  The distribution for the energy flux is symmetric, unlike that
     for the energy density. In both cases, the distribution may possess an integrable singularity. The functional form of the flux distribution function involves
     a modified Bessel function, and is distinct from the shifted Gamma form for the energy density. 
     By considering the joint distribution of energy flux and energy density, we show that the distribution of energy flux tends to be more centrally concentrated than that of the energy density. We also determine the distribution of energy fluxes, conditioned on the energy density being negative. Some applications of the results will be discussed.  
     \end{abstract}

\maketitle

\baselineskip=14pt

\section{Introduction}
\label{sec:intro}

This paper will deal with some exact solutions for the vacuum state probability distribution for the fluctuations of the energy flux in two
dimensional conformal quantum field theories. The vacuum expectation value of the flux vanishes, but an individual measurement can return a nonzero
result, which is equally likely to be positive or negative. Furthermore, the probability of finding a given magnitude for the flux is independent of
its sign, so the distribution will be symmetric. This work will extend previous exact results for the energy density in two 
dimensions~\cite{FFR2010,FH2019,AF2020,Fe&Ho05}, and is related to approximate results in four dimensions~\cite{FF15,FFR2012,FF2020,SFF18,WSF21,WSF23}.  
The flux operator must be averaged in time to be well defined, and the averaging function
describes the details of a physical measurement.  Here we consider two classes of averaging functions which have previously been used
for the energy density. These include the Gaussian and Lorentzian functions as special cases.
We also consider the joint distribution of energy flux and energy density and, by computing the probability that the absolute value of the flux is less than the absolute value of energy density, show that the energy flux is the typically more centrally concentrated of the two. Furthermore, we show that the distribution of fluxes changes markedly when conditioned on the energy density being negative, despite the fact that the probability for obtaining a negative energy density is typically greater than $1/2$.
Some applications of the results to four dimensional models, numerical simulations of quantum fluctuations, and analog models in condensed
matter systems will be discussed. 

\section{Vacuum Energy Density}
\label{sec:energy}

Here we review results from Refs.~\cite{FFR2010,FH2019,AF2020}  for the probability distributions for the energy density in (unitary, positive energy) conformal field theories (CFTs) including the massless free scalar field as a special case. 
The stress tensor for such a theory is determined by  right-moving and left-moving components, $T_R(u)$ and $T_L(v)$, which have the same central charge assuming parity invariance. Here $u = t-x$ and $v =t+x$
are null coordinates on Minkowski spacetime.  These components have a well-defined probability distribution only if they have been averaged by a suitable sampling function,
$f(u)$ or $f(v)$, conveniently normalized so that 
\begin{equation}
 \int_{-\infty}^\infty f(u)\, du = 1\,.
 \end{equation}
Let $\omega_R$ denote the dimensionless averaged operator
\begin{equation}
\omega_R =  \tau^2 \, \bar{T}_R = \tau^2 \,  \int_{-\infty}^\infty f(u) \,  T_R(u)\, du\,,
 \end{equation}
and define $\omega_L$ and $\bar{T}_L$ similarly in terms of $T_L(v)$. Here $\tau$ is a timescale, for example a characteristic width of the sampling function. 
We are interested in the probability distribution $P_{L,R}(\omega)$ for the outcomes of individual measurements of $\omega_{L,R}$ in the vacuum state of the theory.
Explicit results~\cite{FFR2010,FH2019,AF2020} have been
found for specific choices of $f$, including generalizations of the Gaussian and Lorentzian functions. In all these cases, the probability distributions are of the form
of a shifted Gamma distribution,
 \begin{equation}
	P_{L,R}(\omega)= \frac{\beta^\alpha}{\Gamma(\alpha)}\theta(\omega+\omega_0) e^{-\beta(\omega+\omega_0)}(\omega+\omega_0)^{\alpha-1}
\label{eq:PLR}	
\end{equation}
with dimensionless  positive parameters $\omega_0$, $\alpha$, and $\beta$ which depend upon the choice of $f(u)$ and the central charge of the CFT. Note that our notation differs slightly from
that of Refs.~\cite{FFR2010,FH2019,AF2020}, where $\omega$ and $\omega_0$ have dimensions of $length^{-2}$ and $\beta$ has 
dimensions of $length^{2}$.
Features of note are that $P_{L,R}$ vanishes for $\omega<-\omega_0$, which is also the quantum inequality bound for a null component of energy density~\cite{Fe&Ho05,FFR2010}, and displays an integrable singularity as $\omega\to -\omega_0+$ if $\alpha<1$. The probability of obtaining a negative result, $\int_{-\omega_0}^0 P_{L,R}(\omega)d\omega$ is (often substantially) greater than the probability 
$\int_0^\infty P_{L,R}(\omega)d\omega$ of a positive result, but the mean of the distribution is zero, in agreement with the expectation of $\omega_{L,R}$ in the vacuum state.

The energy density is the sum of the  right-moving and left-moving components: 
\begin{equation}
 \rho = T_R + T_L\,.
\end{equation}
Because these components are decoupled from one another (see Section~\ref{sec:joint}) and have the same central charge, their vacuum fluctuations are identical and independent, and the energy density probability distribution is
a convolution of $P_R$ and $P_L$:
\begin{equation}\label{eq:Prho}
 P_{\rho}(x)=\int_{-\infty}^\infty d\omega\, P_L(\omega)\, P_R( x-\omega) \, .
 \end{equation}
 Here $x = \rho \, \tau^2$ is the dimensionless averaged energy density.
Explicit evaluation of the above integral using Eq.~\eqref{eq:PLR}  leads to
\begin{equation}
  P_{\rho}(x) = \frac{\beta^{2\alpha}}{\Gamma(2 \alpha)} \,\theta(x + 2 \omega_0) 
  {\rm e}^{-\beta(x + 2 \omega_0)} \, (x+ 2 \omega_0)^{2 \alpha-1} \,.
 \end{equation}
In effect, the parameters $\omega_0$ and $\alpha$ have been doubled, reflecting the fact that these are proportional to the central charge of the conformal
field theory. 

Note that $P_{\rho}(x) = 0$ for $x < - 2 \omega_0$, the lower bound on the allowed averaged energy density in the vacuum state~\cite{Fe&Ho05}. Also note that $P_{\rho}(x)$ decays exponentially for large $x$, and displays an integrable singularity at the lower bound $-2\omega_0$ 
if $0< \alpha < 1/2$.
Although this result is specific to two dimensional spacetime, some of its features, such as the existence of a lower bound on the energy density
probability distribution, also apply in four dimensions. More generally, the explicit two dimensional results can be a guide to possible effects
in four dimensions. For example, they were used by Carlip {\it et al}~\cite{CMP11,CMP18} to explore effects near a cosmological singularity.

\section{Vacuum Energy Flux}
\label{sec:flux}

The dimensionless averaged energy flux operator is determined by an off-diagonal component of the averaged stress tensor
\begin{equation}
 F = \tau^2 \, \bar{T}^{tx} =   \tau^2 \, (\bar{T}_R - \bar{T}_L ) \,.
 \end{equation}
The probability distribution for vacuum energy flux fluctuation may be constructed as a convolution, in analogy with that for the energy density:
\begin{equation}
	P_{F}(F)= \int_{-\infty}^\infty d\omega \, P_L(\omega)\, P_R(- F+\omega)
	                        =\int_{-\infty}^\infty d\omega \, P_L(\omega+ F/2)\, P_R(\omega - F/2) \,.
\end{equation}
This is an even function of $F$,  as follows from physical grounds and the fact that $P_R = P_L$.
We now proceed to use Eq.~\eqref{eq:PLR} to find it explicitly  in cases where $P_L$ and $P_R$ are shifted Gamma distributions:
\begin{align}
P_{F}(F) &= \frac{\beta^{2\alpha} e^{-2\beta\omega_0}} {\Gamma(\alpha)^2} \int_{-\infty}^\infty d\omega \, 
\theta(\omega + F/2+\omega_0) \theta(\omega - F/2+\omega_0) e^{-2\beta \, \omega}
\;[ (\omega +\omega_0)^2 - F^2/4 ]^{\alpha-1} \nonumber\\
&= \frac{\beta^{2\alpha}e^{-2\beta\omega_0}}{\Gamma(\alpha)^2} \int_{|F|/2 - \omega_0}^\infty d\omega \, 
e^{-2\beta\omega}\;
[ (\omega +\omega_0)^2 - F^2/4]^{\alpha-1}   \nonumber\\
&= \frac{\beta^{2\alpha}}{\Gamma(\alpha)^2}  \int_{|F|/2}^\infty d\omega \, 
e^{-2\beta\omega}\;
 (\omega^2 - F^2/4)^{\alpha-1}     \nonumber\\
&= \frac{\beta^{2\alpha}}{\Gamma(\alpha)^2}  (|F|/2)^{2 \alpha -1}    \int_{1}^\infty d\eta \, 
e^{- \beta |F| \,\eta} \;
(\eta^2- 1)^{\alpha-1}  \,.  
\label{eq:PF0}
\end{align}
Note that $P_{F}(F)$ is independent of the parameter $\omega_0$. Using standard formulae (see 3.387.3 in~\cite{GR} or 10.32.8 in~\cite{NIST}) we may evaluate the integral as a modified Bessel function to find 
\begin{equation}
P_{F}(F) = \frac{ \beta\, (\beta |F|/2)^{\alpha-1/2}}{\sqrt{\pi}\, \Gamma(\alpha) } K_{\alpha-1/2}(\beta |F|) \,.
\label{eq:PF}
\end{equation}
Here the constants $\alpha$ and $\beta$ depend upon the specific choice of sampling function
and the assumption that $P_L=P_R$ are shifted Gamma distributions.
The distribution $P_F$ is a centered symmetric variance-gamma distribution in which $\alpha$ is the shape parameter and $\beta$ is an inverse width scale; in fact, it has been known for a long time that the difference of two identically distributed Gamma distributions is distributed in this way~\cite{Kullback:1936}, and the same applies to the shifted Gamma case.

The asymptotic probability distribution for large argument is
 \begin{equation}
	P_{F}(F) \sim \frac{\beta}{2\Gamma(\alpha)}\left(\frac{\beta |F|}{2}\right)^{\alpha-1}  \, {\rm e}^{- \beta\, |F|} \, [1 +O((\beta |F|)^{-1} )] \,,
	\qquad \beta \, |F| \gg 1 \,.
	\label{eq:asymptotic}
\end{equation}
Thus, the probability of  a large fluctuation decreases exponentially with the same decay constant as for $P_\rho$.

Near the origin,  we can use the asymptotic formula
\begin{equation}
(z/2)^\nu K_\nu(z)\sim \begin{cases} \tfrac{1}{2}\Gamma(\nu) & \nu>0  \\
	-\log(z/2) &\nu=0\\
	  \tfrac{1}{2}\Gamma(-\nu)  (z/2)^{2\nu} & \nu<0, 
	  \end{cases}
\end{equation}
to see that the distribution satisfies
\begin{equation}
		P_{F}(F) \sim \frac{\beta (\beta|F|/2)^{2 \alpha-1} \, \Gamma( 1/2-\alpha)}{2\sqrt{\pi}\, \Gamma(\alpha)  } 
		\label{eq:smallF-1}
\end{equation}
for $\alpha<1/2$,
\begin{equation}
	P_{F}(F) \sim \frac{\beta}{\pi}\log (\beta|F|/2)
	\label{eq:smallF-2}
\end{equation}
for $\alpha=1/2$, 
and 
\begin{equation}
	P_F(F)\sim \frac{\beta\Gamma(\alpha-1/2)}{2\sqrt{\pi}\Gamma(\alpha)}
	\label{eq:smallF-3}
\end{equation}
for $\alpha>1/2$. In particular, $P_F$ has an integrable singularity at $F=0$ for $\alpha\le 1/2$,
and is continuous for $\alpha>1/2$.

The variance of the flux fluctuations is
\begin{equation}
 \langle F^2 \rangle = \int_{-\infty}^\infty dF \, F^2\, P_{F}(F )= \frac{2 \, \alpha }{\beta^2} \,,
\end{equation}
as can be seen using the integral identity  6.561.16 in~\cite{GR} or 10.43.19 in~\cite{NIST}.
It is also of interest to examine the cumulative distribution function
\begin{equation}
	P_<(F) =  \int_{-\infty}^F dy \, P_F(y) \,,
\end{equation}
which is the probability to find a value less than $F$ in a measurement. It may be calculated in terms of modified Bessel and Struve functions as
\begin{equation}
P_<(F) = \frac{1}{2}  \{ 1 + \beta F [ K_{\alpha-1/2}(\beta |F|)\, L_{\alpha-3/2}(\beta |F|) + K_{\alpha-3/2}(\beta |F|)\, L_{\alpha-1/2}(\beta |F|) ] \} \,.
\label{eq:PCE}
	\end{equation}
Here we have again used	 the identities 6.561.16, as well as 6.561.4, in Ref.~\cite{GR} (see also 10.43.2 in Ref.~\cite{NIST}).  We may use either the above result, or numerical integration of
Eq.~\eqref{eq:PF}, to obtain equivalent numerical results for  $P_<(F)$.

\section{The joint distribution of averaged flux and energy density}\label{sec:joint}

The results of Sections~\ref{sec:energy} and~\ref{sec:flux} can be understood from a more general perspective that leads to further results. 
As the two averaged null components $\omega_R$, $\omega_L$ of the stress tensor commute, they have a joint probability distribution, whose joint moment generating function 
	is the product of the two independent generating functions
	\begin{equation}\label{eq:GFfac}
		\langle {\rm e}^{s \omega_R+t \omega_L}\rangle = \langle {\rm e}^{s \omega_R}\rangle
		\langle {\rm e}^{t \omega_L}\rangle ,
	\end{equation}
because the vacuum expectation values factorize (see Appendix~\ref{appx:factorisation})
\begin{equation}
\langle \omega_R^m \omega_L^n\rangle = \langle \omega_R^m \rangle \, \langle \omega_L^n\rangle.
\end{equation}
Consequently, the joint probability distribution function is the product of the two individual ones, giving
\begin{equation}
	\Prob((\omega_R,\omega_L)\in \Delta)= \int_{\Delta} P_{R}(\lambda)P_{L}(\mu)\, d\lambda\,d\mu
\end{equation}
for any Borel subset $\Delta\subset \mathbb{R}^2$.
In the above expressions, the generating functions may be understood as formal series in $s$ and $t$.
Petersen's theorem (see~\cite{Schmuedgen_momentbook}, Theorem 14.6) states that the joint moments uniquely determine the joint probability distribution provided that the marginal distributions are uniquely determined by their moments. This is certainly satisfied if $P_{L,R}$ are given by shifted Gamma distributions, because their moments satisfy the Hamburger moment condition~\cite{FFR2010}.

The averaged flux $F=\omega_R-\omega_L$ and energy density $\rho=\omega_R+\omega_L$ also commute and have a joint probability distribution given in terms of that of $\omega_L$ and $\omega_R$. Specifically,
\begin{equation}
	\Prob((\rho,F)\in \Delta) = \int_{\widetilde{\Delta}} P_{R}(\lambda)P_{L}(\mu)\, d\lambda\,d\mu = \int_{{\Delta}}   P(\rho,F) \,   d\rho\,dF \, ,
\end{equation}
where $\widetilde{\Delta}=\{(\lambda,\mu)\in\mathbb{R}^2: (\lambda+\mu,\lambda-\mu)\in\Delta \}$,  and the joint probability density function is
\begin{equation}
 P(\rho,F) =  \frac{1}{2}\, P_{R}( \tfrac{1}{2}(\rho+F))P_{L}( \tfrac{1}{2}(\rho-F))\, .
 \label{eq:PrhoF}
\end{equation}
 The factor of $\tfrac{1}{2}$ arises from the change of variables. Integrating out $F$ or $\rho$ in~\eqref{eq:PrhoF_Gamma}, one obtains the formulae~\eqref{eq:Prho} and~\eqref{eq:PF} for the marginal distributions of $\rho$ and $F$ as above.
To understand the significance of the joint probability density function, consider the simple example $\Delta = \{\rho_0 \leq \rho \leq \rho_0 + \Delta \rho, F_0 \leq F \leq F_0 + \Delta F \}$, where $\Delta \rho \ll |\rho_0|$ and  $\Delta F \ll |F_0|$, 
in which limit we have  
\begin{equation}
	\Prob((\rho,F)\in \Delta) \approx P(\rho_0,F_0) \,  \Delta \rho \, \Delta F\,.
\end{equation}  

Due to $P_L$ and $P_R$ being supported in $[-\omega_0,\infty)$, $P$ effectively has
a factor of $\theta(\rho + F + 2 \omega_0)\theta(\rho - F  + 2 \omega_0)$ and one easily sees that
$P$ is supported in the set of $(\rho,F)\in\mathbb{R}^2$ such that $|F| \le \rho +  2 \omega_0$. 
As $\rho \rightarrow - 2 \omega_0$, the quantum inequality lower bound, the flux must vanish
in the following sense: for any $\rho_0>-2\omega_0$, the conditional random variable $F|(\rho<\rho_0)$ (i.e., $F$ conditioned on $\rho$ being less than $\rho_0$) takes values in the interval $[-2\omega_0-\rho_0,2\omega_0+\rho_0]$ and therefore the expectation value of its absolute value is bounded above by $2\omega_0+\rho_0$, which vanishes
in the limit $\rho_0\to-2\omega_0$. Thus, in this limit, $F|(\rho<\rho_0)$ converges in mean to the random variable taking the constant value $0$. Note that this conclusion does not depend upon the specific functional forms of $P_R$ and $P_L$. 

In the case where $P_R$ and $P_L$ are identical shifted Gamma distributions, a similar calculation to one given in Eq.~\eqref{eq:PF0} shows that
\begin{equation}\label{eq:PrhoF_Gamma}
	\Prob((\rho,F)\in \Delta) = 
	\frac{2(\beta/2)^{2\alpha}}{\Gamma(\alpha)^2}\int_0^\infty d\rho'\,
	\int_{-\rho'}^{\rho'} dF\, \chi_\Delta(\rho'-2\omega_0,F) {\rm e}^{-\beta\rho'} ((\rho')^2-F^2)^{\alpha-1},
\end{equation}
where $\chi_\Delta$ is the characteristic function of $\Delta$, i.e., $\chi_\Delta(\rho,F)=1$ for $(\rho,F)\in\Delta$ and $\chi_\Delta(\rho,F)=0$ otherwise. 

Using the joint distribution, we can determine, for example, the probability that $|F|<|\rho|$,
corresponding to $\Delta=\{(\rho,F): |F|<|\rho|\}$, for which $\widetilde{\Delta}$ is the union of the first and third open quadrants of $\mathbb{R}^2$. With $P_L=P_R$ (but not necessarily of the shifted Gamma form) this gives 
\begin{equation}
	\Prob(|F|<|\rho|) = p^2 + (1-p)^2 = 1-2p+2p^2
\end{equation}
where $p=\Prob(\omega_{L,R}<0)$. One typically has $p>\tfrac{1}{2}$, because the expectation is zero and the distribution has a long positive tail but
cannot take negative values below the quantum inequality bound. It follows that
$\Prob(|F|<|\rho|) >\tfrac{1}{2}$, as well, indicating the flux tends to be more centrally concentrated than the energy density. This effect can be quite marked: for instance, with $c=1$ and Gaussian smearing, $p=0.89$~\cite{FFR2010} and $\Prob(|F|<|\rho|) = 0.81$ (quoting values to $2$d.p.). 

As another example, we can compute the probability density function $P_{F|\rho<0}(F)$ of the flux, conditioned on the energy density being negative (thus lying in $[-2\omega_0,0]$), which is given in general by 
\begin{equation}
P_{F|\rho<0}(F) = \frac{1}{2q} \int_{-2\omega_0}^0 d\rho\, P_R(\tfrac{1}{2}(\rho+F))P_L(\tfrac{1}{2}(\rho-F)),
\end{equation}
where $q=\Prob(\rho<0)$. This formula is obtained by setting $\Delta=(-2\omega_0,0)\times (-\infty,F)$ in~\eqref{eq:PrhoF} and differentiating with respect to $F$, then dividing by $q$ to normalize the distribution.
 Given that $P_L$ and $P_R$ both have support $[-\omega_0,\infty)$, we see that 
$P_{F|\rho<0}(F) =0$ for $|F|>2\omega_0$, so this distribution has compact support $[-2\omega_0,2\omega_0]$. In the case where $P_{L,R}$ are identical shifted Gamma distributions, this gives 
\begin{align}
	P_{F|\rho<0}(F) &= \frac{\beta^{2\alpha}{\rm e}^{-2\beta\omega_0}}{q\Gamma(\alpha)^2 2^{2\alpha-1}} \int_{-2\omega_0}^0 d\rho \,
	\theta(\rho+F+2\omega_0)	\theta(\rho-F+2\omega_0) {\rm e}^{-\beta \rho}((\rho+2\omega_0)^2-F^2)^{\alpha-1}\nonumber \\
	&=\frac{\beta^{2\alpha}\theta(2\omega_0-|F|)}{q\Gamma(\alpha)^2 2^{2\alpha-1}}
	\int_{|F|}^{2\omega_0} d\rho' \,  {\rm e}^{-\beta \rho'}((\rho')^2-F^2)^{\alpha-1}\nonumber \\
	&=\frac{\beta(\beta |F|/2)^{2\alpha-1}}{q\Gamma(\alpha)^2} Q(\alpha,\beta|F|;|F|/(2\omega_0))\theta(2\omega_0-|F|), 
	\label{eq:F-rho-neg}
\end{align}
where
\begin{equation}
	Q(\alpha,\eta;z) = \int_1^{1/z} ds\, {\rm e}^{-\eta s }(s^2-1)^{\alpha-1}.
\end{equation}

Consider the limit in which $|F|\to 2\omega_0-$, so $1/z = 1 + \epsilon$, with $0<\epsilon = 2\omega_0/|F| -1   \ll 1$, and $\eta = 2\omega_0\beta/(1+\epsilon) \to  2\omega_0 \beta$. Now $Q/\epsilon^{\alpha}$ becomes
\begin{equation}
\epsilon^{-\alpha}Q = \frac{1}{\epsilon^\alpha}\int_0^\epsilon dx \, {\rm e}^{-\eta (1+x) } \,[(1+x)^2-1]^{\alpha-1}  \longrightarrow {\rm e}^{-2\omega_0 \beta} \,   \frac{2^{\alpha -1}}{\alpha} ,
\end{equation}
as can be confirmed by a dominated convergence argument, leading to 
\begin{equation}
P_{F|\rho<0}(F) \approx \frac{\beta^{2\alpha}\, \omega_0^{\alpha-1} \,{\rm e}^{-2\beta\omega_0}}{2 \,q \, \alpha\, \Gamma(\alpha)^2 } \; (2\omega_0 - |F|)^\alpha 
\end{equation}
as $|F|$ approaches $2\omega_0$ from below. An example of this behavior will appear in 
Fig.~\ref{fig:PFrhoneg} below. 

We can also investigate the limit in which $|F|\ll \omega_0$. Here, it is convenient to use the
penultimate expression in~\eqref{eq:F-rho-neg}, also noting that the exponential factor lies in 
the interval $[{\rm e}^{-2\omega_0 \beta},1]$ for all $\rho'\in[0,2\omega_0]$. Thus
\begin{equation}\label{eq:PFrhonegasymp1}
P_{F|\rho<0}(F)	\asymp  I(F) =  \int_{|F|}^{2\omega_0} d\rho' \,  ((\rho')^2-F^2)^{\alpha-1}
\end{equation}
as $|F|\to 0$,
where the $\asymp$ symbol means that the left-hand side is bounded above and below by constant multiples of the right-hand side. There are three cases. For $0<\alpha< \tfrac{1}{2}$,
we have
\begin{equation}
	P_{F|\rho<0}(F)	\asymp F^{2\alpha -1} \int_{1}^{2\omega_0/|F|} ds \,  (s^2-1)^{\alpha-1}
\end{equation}
and because the integral is finite and nonzero in the limit $|F|\to 0$, it follows that
$P_{F|\rho<0}(F)\asymp F^{2\alpha -1}$, which is an integrable singularity (see Fig.~\ref{fig:PFrhoneg} for an example). On the other hand, if $\alpha=\tfrac{1}{2}$, 
the above integral can be evaluated exactly and grows logarithmically in $2\omega_0/|F|$,
so $P_{F|\rho<0}(F)	\asymp \log(2\omega_0/|F|)$. Finally, if $\alpha>\tfrac{1}{2}$, an integration by parts argument based on multiplying and dividing the integrand of Eq.~\eqref{eq:PFrhonegasymp1}
by $2\rho'$ gives
\begin{equation}
	I(F)=\frac{((2\omega_0)^2-F^2)^{\alpha}}{4\omega_0\alpha} + \frac{1}{2\alpha}\int_{|F|}^{2\omega_0} d\rho' \, \frac{1}{(\rho')^2} ((\rho')^2-F^2)^{\alpha},
\end{equation}
where $I(F)$ is the integral on the right-hand side of Eq.~\eqref{eq:PFrhonegasymp1}. As $(\rho')^{-2}< ((\rho')^2-F^2)^{-1}$, the second term on the right-hand side is bounded above by $I(F)/(2\alpha)$. On rearranging,
\begin{equation}
	I(F)< \frac{2\alpha}{2\alpha-1}\frac{((2\omega_0)^2-F^2)^{\alpha}}{4\omega_0\alpha} ,
\end{equation}
which shows that $P_{F|\rho<0}(F)$ remains bounded as $|F|\to 0$ for $\alpha>\tfrac{1}{2}$. 
Note that the behavior we have given here results in the same singularity structure exhibited
by $P_F(F)$ as $|F|\to 0$, described in Eqs.~\eqref{eq:smallF-1}, \eqref{eq:smallF-2}, and \eqref{eq:smallF-3}.

\section{Specific Sampling Functions}
 
  Here we discuss some choices of sampling function for which the probability $P_{L,R}$ distributions take the shifted Gamma form given in Eq.~\eqref{eq:PLR}.
  
  \subsection{Gaussian-like Functions}
  
The class of functions of the form
 \begin{equation}
 f_{a,b}(u) = \gamma \, u^{2 a}\,  {\rm e}^{-b\, u^2}\, ,
 \label{eq:gen-gaussian}
\end{equation}
where $a$ is a nonnegative integer, $b > 0$  and $\gamma = b^{a + 1/2}/\Gamma(a + 1/2)$,  was discussed in Ref.~\cite{AF2020}.
Here we may take
\begin{equation}
 \tau = \frac{1}{\sqrt{b}} \,.
\end{equation}
The results of Ref.~\cite{AF2020} may be expressed in our present notation as
\begin{equation}
 \alpha = \frac{c\, ( 4 a -1)}{ 24 (2a -1)}
 \label{eq:alpha-gen-gaussian}
\end{equation}
and
\begin{equation}
 \beta = \pi \,.
\end{equation}
Here $c$ is the central charge of the conformal field theory, and $c = 1$ for a massless scalar field. The usual Gaussian, which was
treated in Ref.~\cite{FFR2010}, is the case $a = 0$, so
\begin{equation}
\alpha = \frac{c}{24} \,, \qquad \beta = \pi \,. 
\end{equation}

\begin{figure}
\includegraphics[width=0.6\linewidth]{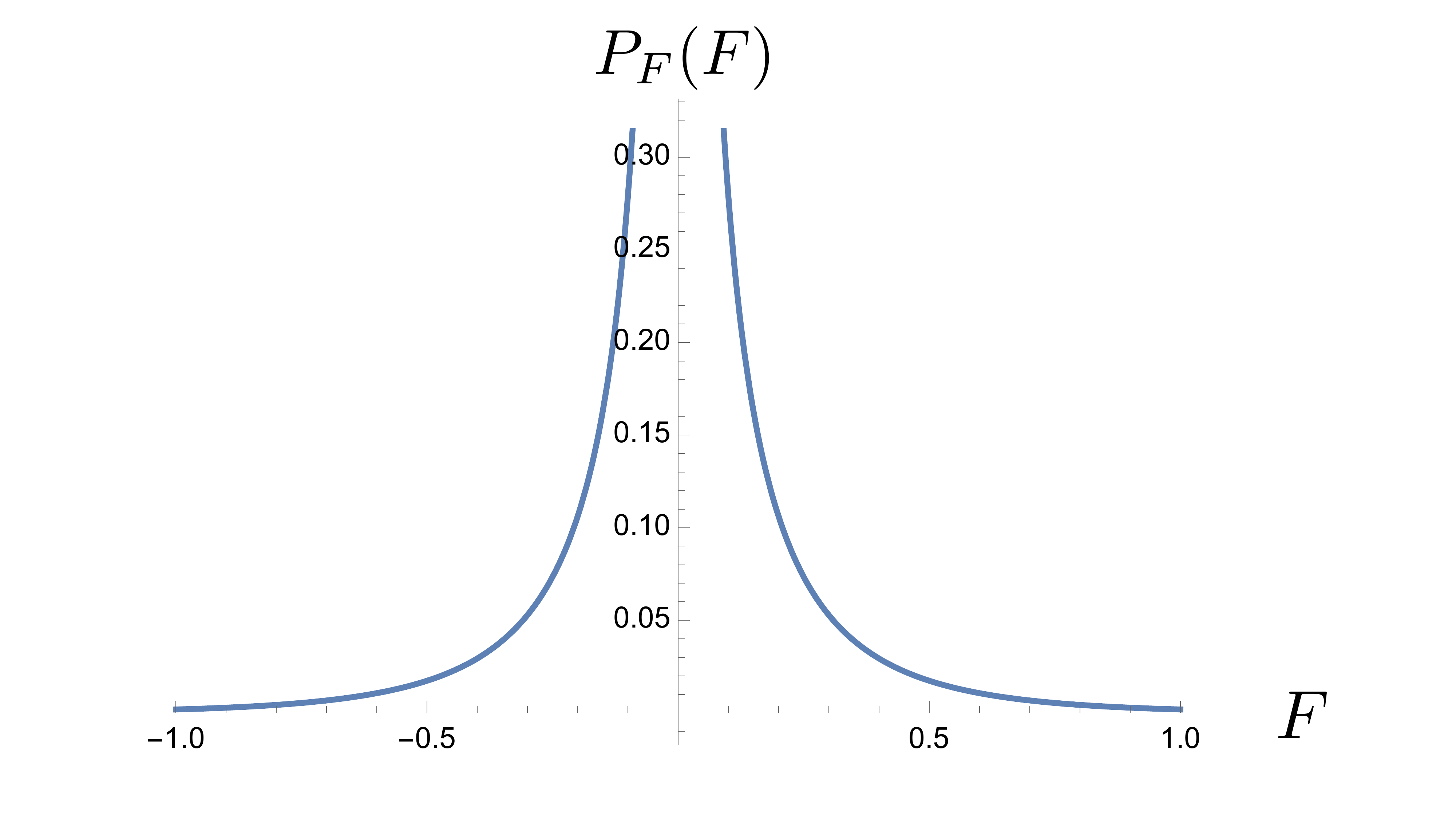}
\caption{The probability distribution, $P_F(F)$, with Gaussian averaging is plotted as a function of the dimensionless flux, $F$.}
\label{fig:PG}
\end{figure}

The vacuum probability distribution  for the Gaussian averaged flux of a massless scalar field, $c = 1$,
is plotted in Fig~\ref{fig:PG}.    Here $P_F \propto |F|^{-11/12}$ near the origin.

\begin{figure}
    \centering
    \begin{minipage}[c]{\linewidth}
    \includegraphics[width=0.6\linewidth]{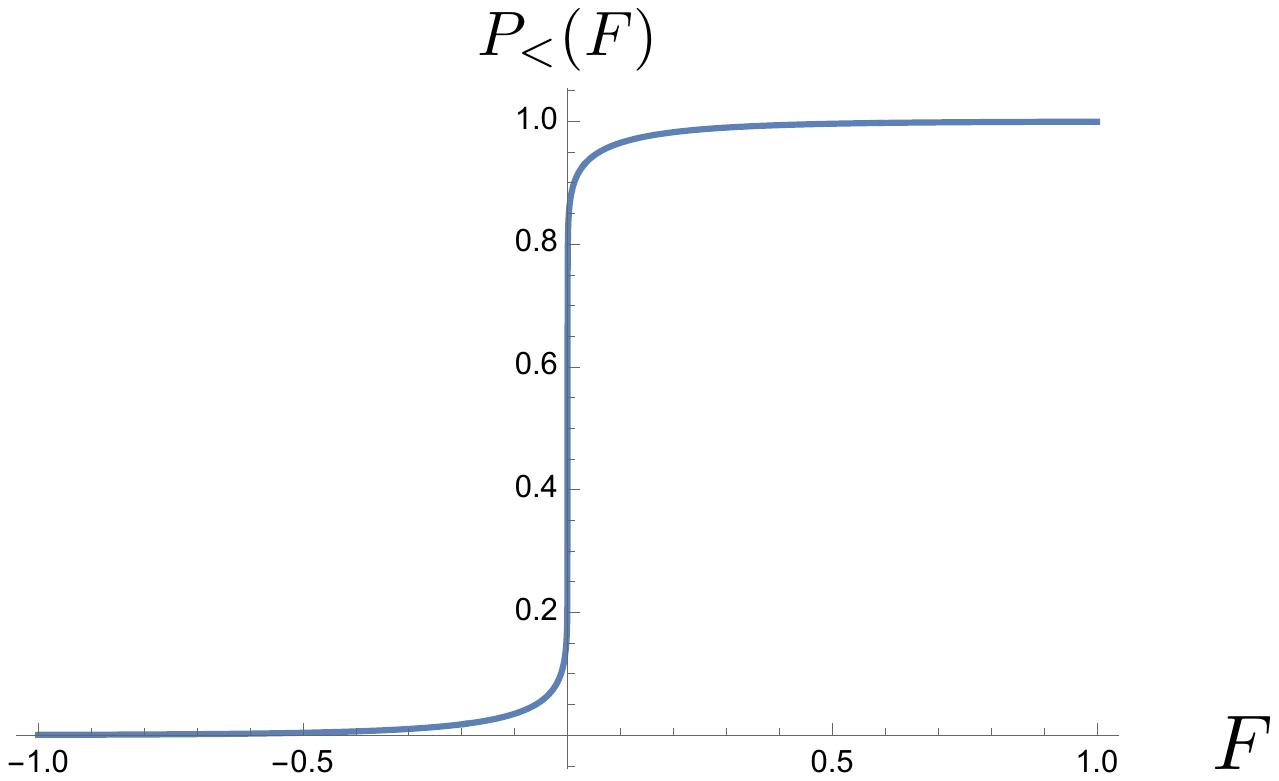}
    \end{minipage}
   \\
    \vspace{2mm}
    \begin{minipage}[c]{\linewidth}
    \includegraphics[width=0.6\linewidth]{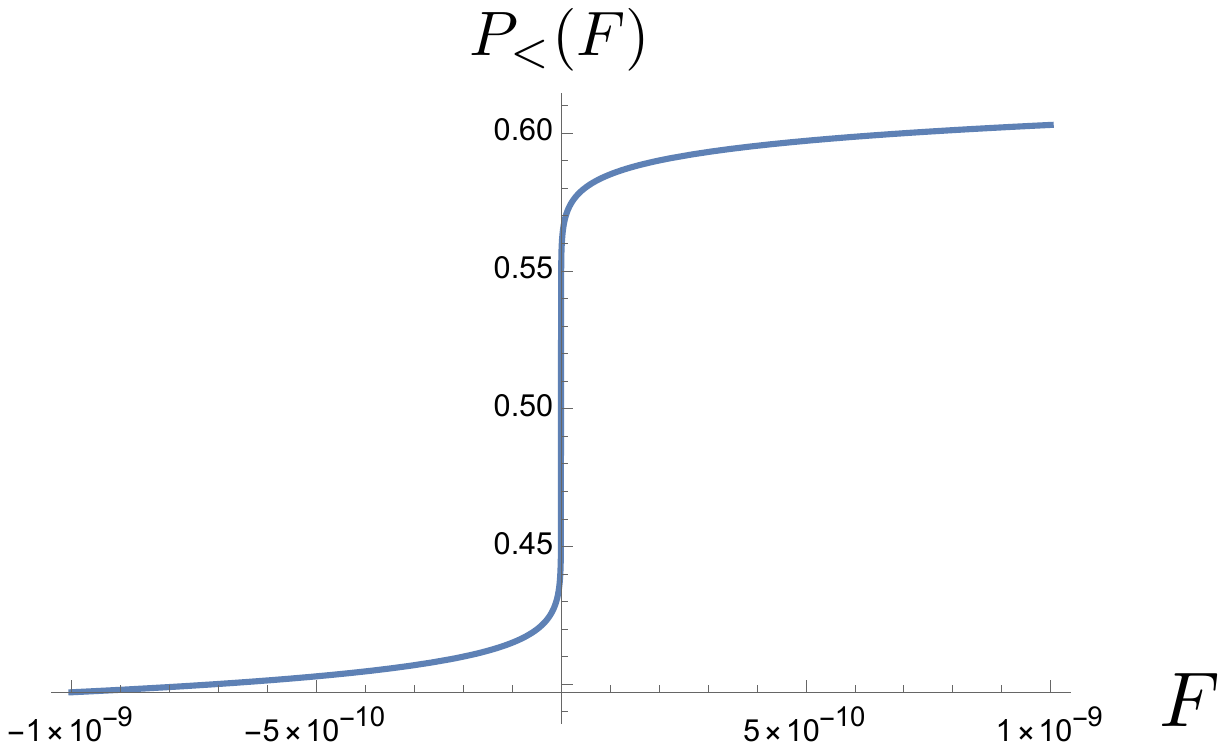}
    \end{minipage}
\caption{The cumulative probability distribution, $P_<(F)$, with Gaussian averaging is plotted as a function of the dimensionless flux, $F$, on two different scales.
These plots may be obtained by numerical integration of Eq.~\eqref{eq:PF} or numerical evaluation of the exact expression, Eq.~\eqref{eq:PCE}.} 
\label{fig:PGC}
\end{figure}

The cumulative distribution function $P_<(F)$ 
is plotted for the Gaussian averaged scalar field case with $c=1$  in Fig.~\ref{fig:PGC}.  Note that the integrable singularity in  $P_F(F) $ leads to $P_<(F)$
approximating a step function. Numerical evaluation gives $P_<(10^{-10})=0.585$ and $P_<(-10^{-10})=0.415$, so the average slope in the interval $|F| < 10^{-10}$ is
	\begin{equation}
		\frac{\Delta P_<}{\Delta F} \approx  8.5 \times 10^8 \,.
	\end{equation}
	which is consistent with the lower panel in Fig.~\ref{fig:PGC}.

\begin{figure}
	\includegraphics[width=0.5\textwidth]{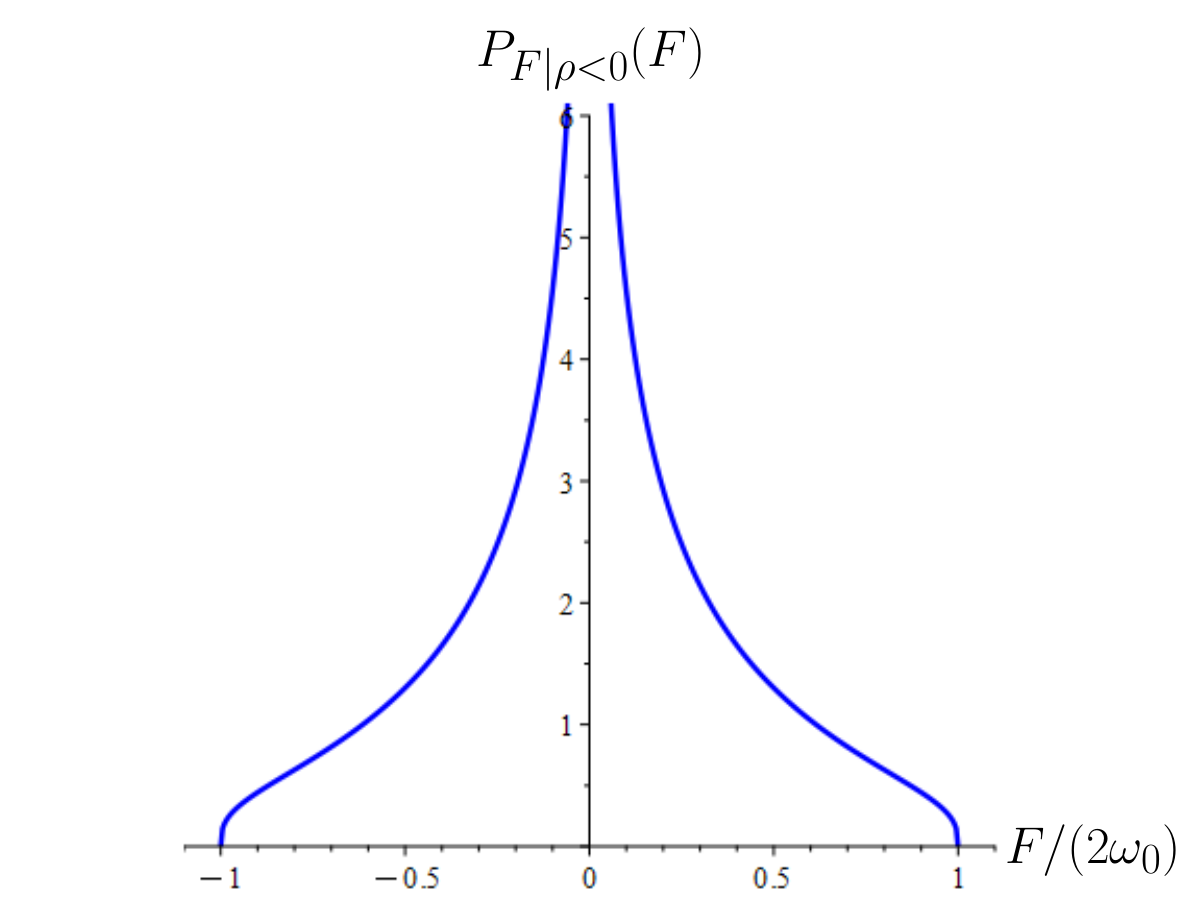}
	\caption{The probability distribution of flux, conditioned on the energy density being negative.}
	\label{fig:PFrhoneg}
\end{figure}

The distribution $P_{F|\rho<0}(F) $ given in Eq.~\eqref{eq:F-rho-neg} is plotted in Fig.~\ref{fig:PFrhoneg} for values $\alpha=c/24$, $\beta=\pi$, 
with central charge $c=8$ to indicate better the shape of the plot near the end of the support. 

  \subsection{Lorentzian-like Functions}
 
  Another class of functions considered in Ref.~\cite{AF2020} takes the form
 \begin{equation}
 g_{n,a,b}(u)   = \frac{C \, u^{2 a}}{(b^2 + u^2)^n} \,, 
\end{equation}
where $ 0 \leq a < n$ are integers and $b>0$, and we may now take
 \begin{equation}
 \tau = b\,.
 \end{equation}
 Now
\begin{equation}
 \alpha = \frac{c n ( 4a^2 -4 a n - 4 a +n)}{12 ( 2a -2n -1)(n+1)(2a-1)}\,,
\end{equation}
and
\begin{equation}
 \beta = \frac{ 4 n \pi }{4 (n-a)^2 -1} \, . 
\end{equation}

\begin{figure}
\includegraphics[width=0.6\linewidth]{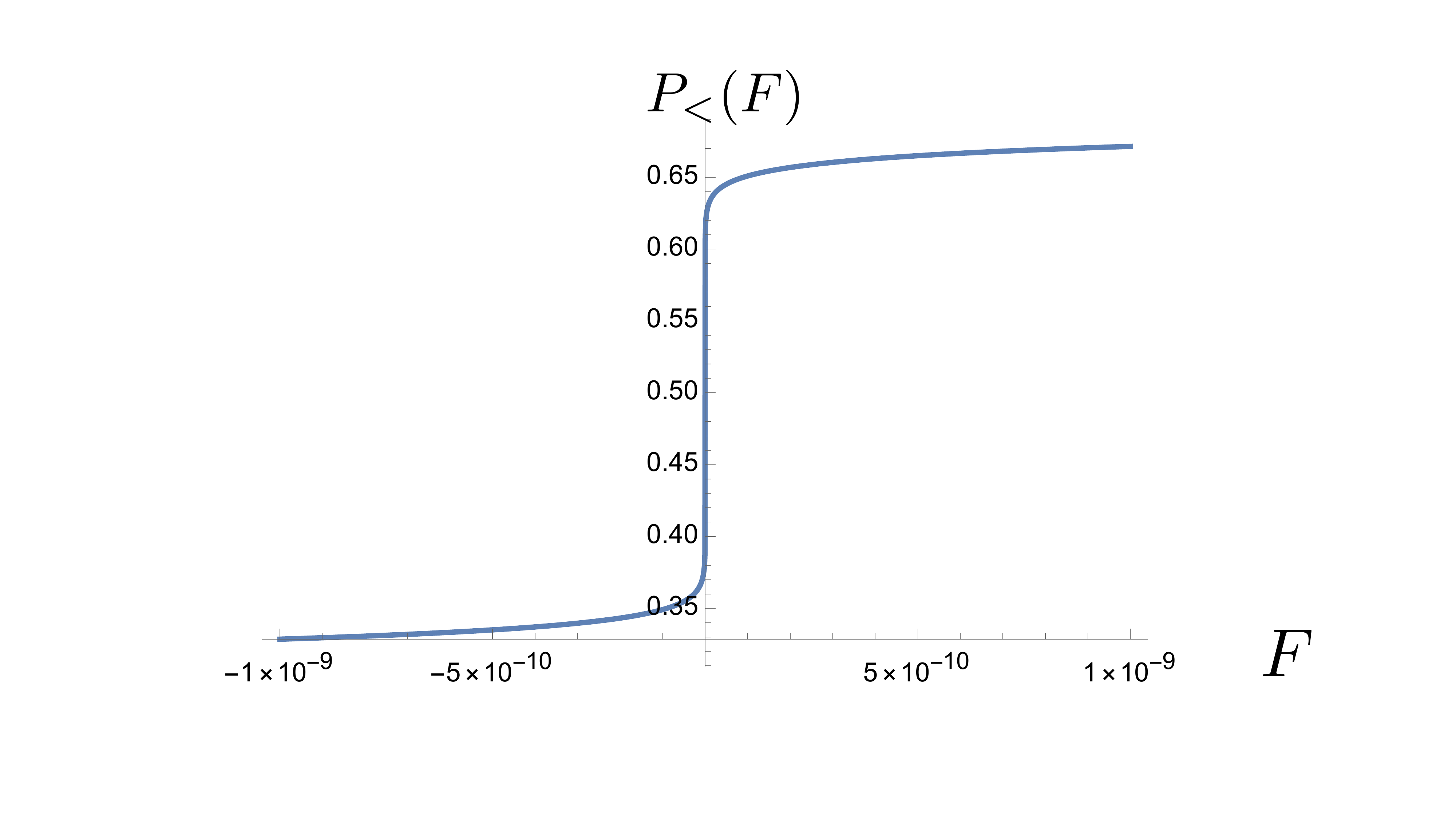}
\caption{The cumulative probability distribution, $P_<(F)$, with Lorentzian averaging is plotted as a function of the dimensionless flux, $F$.}
\label{fig:PLC}
\end{figure}

The usual Lorentzian is
\begin{equation}
 g_{1,0,b}(u) = \frac{b}{\pi(u^2 +b^2)}\,,
\end{equation}
 for which 
\begin{equation}
 \alpha = \frac{c}{72}\,, \qquad \beta = \frac{4 \pi}{3} \, ,
\end{equation}
In the massless scalar field case, $ c = 1$, $P_F(\omega)$ is qualitatively similar to the Gaussian averaged case plotted in Fig.~\ref{fig:PG}. 
 However,  here $P_F \propto |\omega|^{-17/18}$ near the origin, and $P_<(\omega)$ is plotted in Fig.~\ref{fig:PLC}.

We find in this case that $P_<(10^{-10})=0.775$ and $P_<(-10^{-10})=0.2255$, so
 the average slope in the interval $|F| < 10^{-10}$ is of order
\begin{equation}
 \frac{\Delta P_<}{\Delta F} \approx 2.7 \times 10^9 \,
\end{equation}
This is about a factor of $3$ larger than the Gaussian case, and is presumably due to the more singular behavior of  $P_F$ for the Lorentzian case. 

Note that for the massless scalar field case, $ c = 1$, both the Gaussian and Lorentzian averaging functions lead to 
$\alpha <  \frac{1}{2}$, and hence an integrable singularity in $P_F$. This need not be the case for some of the generalized
averaging functions discussed above. For example, if we set $a = \frac{23}{44}$ in 
Eqs.~\eqref{eq:gen-gaussian} and \eqref{eq:alpha-gen-gaussian}, we obtain a generalized Gaussian for which $\alpha =1$,
leading a probability distribution which is finite everywhere. This is a good illustration of how sensitive stress tensor probability distributions are to the details of the measurement process.

 \subsection{Compactly Supported Functions}
 Compactly supported functions, those which vanish outside of finite intervals, are the appropriate descriptions foe physical measures of finite duration.
  All of the smearing functions discussed above have tails, and hence are not compactly supported. A class of infinitely differentiable
 compactly supported functions was described in Ref.~\cite{FF15}, and used in Refs.~\cite{SFF18,WSF21,WSF23}. 
 They have Fourier transforms which decay as an exponential of a fractional power
 of  frequency:
 \begin{equation}
 \hat{f}(\omega) \sim {\rm e}^{-a |\omega|^{\alpha_p}} \, , \qquad 0 < \alpha_p < 1 \,.
\end{equation}
Although the Lorentzian function is not compactly supported, its Fourier transform corresponds to the $\alpha_p =1$ limit of this class.

The focus of Ref.~\cite{FF15} was on the rate of growth of moments and the asymptotic probability distributions for stress tensor components
in four-dimensional spacetime. The explicit example used was that of $: \dot{\varphi}^2:$, the normal ordered square of  the time derivative of a
massless scalar field. This operator appears as part of the energy density and other stress tensor components, and its asymptotic probability distribution
was assume to model those of a general component.  If we set $p=1$ in the discussion in
Sect.~IV of Ref.~\cite{FF15}, we can infer the asymptotic probability distribution for $: \dot{\varphi}^2:$ in two dimensions sampled by a 
compactly supported function. If we assume that asymptotic  distribution also holds for the flux in two dimensions, we have 
\begin{equation}
 P(F) \sim {\rm e}^{- \beta\,  |F|^{\alpha_p}} 
\end{equation}
for some constant $\beta$. The $\alpha_p =1$ limit of this expression agrees with the form found in Eq.~\eqref{eq:asymptotic}.

 \section{Summary}
 \label{sec:sum}

We have treated the probability distributions for vacuum fluctuations of the energy flux in two dimensional spacetime. 
 These distributions  depend upon the details of the sampling function used to find spacetime averages of the flux 
 operator. In many, but not all cases, the distribution has an integrable singularity at the origin. In particular, if 
 $\alpha > \frac{1}{2}$, the distribution $P(F)$ is finite at $F=0$. In all cases, $P(F)$ decays as an exponential 
 for large arguments, an illustration of the non-Gaussian character of vacuum stress tensor fluctuations.
 
 In addition, in Sect.~\ref{sec:joint}, we construct a joint probability distribution for the energy flux and energy density
 operators which have been averaged with the same sampling function. The joint distribution may be used, for example,
 to compute a modified  probability distribution for the energy flux under some condition on the energy density, such as
 being negative.
 
 Both of these sets of results have potential physical applications to four dimensional models and to condensed matter
 systems with effectively one spatial dimension.  Two dimensional models have often been used to infer insights into
 possible behavior in four dimensions~\cite{CMP11,CMP18}.  Electromagnetic vacuum flux fluctuations could  have observable 
 effects on the motion of electrons~\cite{LF24}, and the two dimensional flux fluctuation models may be useful in further
 studies of this effect.
 
 Zero point density fluctuations in a fluid~\cite{WF20} are an analog model for quantum stress tensor fluctuations. 
Thus, phonons in a one space dimensional system form an analog model for two spacetime dimensional quantum field theories, 
and one which may be experimentally accessible.
  
  The results of this paper are expected to be useful for future numerical simulations  of stress tensor fluctuations. 
  Simulations of energy density fluctuations  without correlations were performed in Refs.~\cite{CMP11,CMP18}, and simulations of Gaussian field fluctuations including correlations between  different times were treated in Ref.~\cite{TYF24}.  The results of
  Sect.~\ref{sec:flux} for $P(F)$ will be useful for simulations of energy flux fluctuations which include correlations between
  different times.
  
  The first step in a numerical simulation is an algorithm to generate a set of
  outcomes which obey a given probability distribution. This is usually done using a cumulative probability distribution, such as
  $P_<(F)$. The numerical challenges are greater when $P(F)$ has an integrable singularity, and hence $P_<(F)$ is close to a 
  step function. The non-singular $\alpha > \frac{1}{2}$ cases seem likely to lead to more stable simulations, but all values of
  $\alpha$ are of interest.  A first step will be a simulation of flux fluctuations at different times using both our results for $P(F)$
  and the flux-flux correlation function.
  
  Future simulations of both the energy flux and density fluctuations could employ either a correlation function or a joint
  probability distribution. A preliminary study using a flux-density correlation function was done in Sect.~III of Ref.~\cite{FR07},
  where the correlation between energy density and flux in adjacent space time regions was discussed. An extension of
  this work which uses the detailed probability distributions treated in the present paper is now possible. The joint distribution
  also provides information about correlations about flux and density fluctuations, but when averaged over the same space time
  region. The possibility of a joint distribution for flux and energy density averaged over different regions is an open question.

\begin{acknowledgments} 
This work was supported in part  by the National Science Foundation under Grant PHY-2207903
and EPSRC grant EP/Y000099/1. CJF also thanks Henning Bostelmann and Gandalf Lechner for a useful conversation concerning Appendix~\ref{appx:factorisation}
\end{acknowledgments}

\appendix
\section{Factorization of the joint moment generating function}\label{appx:factorisation}

The decoupling of the left- and right-moving expectation values is well-known in CFT, but we did not find a simple direct argument in the literature. However an argument is easily given, assuming that the CFT obeys standard properties of a QFT, specifically the translational invariance of the vacuum and the cluster property~\cite{StreaterWightman} (more directly, but less simply, one could argue from the spectral properties of the translation operator, cf.~\cite{Maison:1968,Petrov:1973}). Suppose $Z$ is any product of $m$ factors of $\omega_R$ and $n$ factors of $\omega_L$. Because they commute, one has
\begin{equation}
	\langle Z\rangle = \langle \omega_R^m \omega_L^n\rangle = \langle U(\lambda,\lambda)\omega_R^m \omega_L^nU(-\lambda,\lambda)\rangle
\end{equation}
for any $\lambda\in\mathbb{R}$, where $U(t,x)$ is the unitary operator implementing translation through $(t,x)$, and we have used the translation-invariance of the vacuum. But as translation through $(\lambda,\lambda)$ (resp. $(-\lambda,\lambda)$) leaves $u$ (resp., $v$) unchanged, $[U(\lambda,\lambda),\omega_R]=0=[U(-\lambda,\lambda),\omega_L]$. Consequently,
\begin{equation}
	\langle Z\rangle = \langle  \omega_R^m U(0,2\lambda)\omega_L^n \rangle = \langle  \omega_R^m U(0,2\lambda)\omega_L^n U(0,2\lambda)^{-1}\rangle  
\end{equation}
using translation invariance of the vacuum again. As $\lambda\to\infty$, the left-hand side is constant, but the right-hand side tends to $\langle \omega_R^m \rangle\langle\omega_L^n\rangle$ by the cluster property for large spacelike separations, giving $\langle Z\rangle =  \langle \omega_R^m \omega_L^n\rangle$. Thus, for any integer $N\ge 0$,
\begin{equation}
\frac{1}{N!}	\langle (s\omega_R+t\omega_L)^N\rangle = \sum_{\substack{m,n\ge 0\\ m+n=N}}\frac{s^mt^n}{m!n!}\langle \omega_R^m\rangle \langle \omega_L^n\rangle,
\end{equation}
and Eq.~\eqref{eq:GFfac} for the joint moment generating functions follows, understood as 
an equality of formal power series in $s$ and $t$.

\end{document}